\documentclass[aps,prb,twocolumn,superscriptaddress,showpacs,amsmath,amssymb]{revtex4}
\usepackage{graphicx}
\usepackage{dcolumn}
\usepackage{bm}
\usepackage{longtable}
\usepackage{dcolumn}

\begin{document}

\newcommand{\Tc}{T$_{\mathrm C}$\space}
\newcommand{\Tco}{T$_{\mathrm CO}$\space}
\newcommand{\sinth}{\mbox{$\sin\theta/\lambda$}}
\newcommand{\inA}{\mbox{\AA$^{-1}$}}
\newcommand{\mub}{\mbox{$\mu_{B}$}}
\newcommand{\mns}{$-$}
\newcommand{\ddd}{3\textit{d}}
\newcommand{\pp}{2\textit{p}}
\newlength{\minusspace}
\settowidth{\minusspace}{$-$}
\newcommand{\msp}{\hspace*{\minusspace}}
\newlength{\zerospace}
\settowidth{\zerospace}{$0$}
\newcommand{\zsp}{\hspace*{\zerospace}}
\newcommand{\lasrmn}{La$_{2-2x}$Sr$_{1+2x}$Mn$_{2}$O$_7$\space}
\newcommand{\mnt}{Mn$^{3+}$}
\newcommand{\mnf}{Mn$^{4+}$}
\newcommand{\eg}{$\textit{e}_{g}$ }
\newcommand{\tg}{$\textit{t}_{2g}$ }
\newcommand{\qce}{$\textit{q}_{CE}=(1/4,1/4,0)$}
\newcommand{\qcel}{$\textit{q}_{CE}=(1/4,-1/4,0)$}
\newcommand{\qos}{$\textit{q}_{L}=(0.3,0,1)$}
\newcommand{\sqw}{S(\textit{Q},$\omega$)}
\newcommand{\Ts}{T$^{*}$\space}
\newcommand{\GG}{$\Gamma$\space}

\title{Inhomogeneous Magnetism in La-doped CaMnO$_{3}$. (II) Mesoscopic Phase Separation due to Lattice-coupled FM Interactions}

\author{C.D. Ling}
\email{ling@ill.fr}
\affiliation{Institut Laue-Langevin, BP 156, 38042 Grenoble Cedex 9, France}
\affiliation{Materials Science Division, Argonne National Laboratory, Argonne, Illinois 60439}
\author{E. Granado}
\affiliation{NIST Center for Neutron Research, National Institute of Standards and Technology, Gaithersburg, Maryland 20899}
\affiliation{Center for Superconductivity Research, University of Maryland, College Park, Maryland 20742}
\affiliation{Laborat\'{o}rio Nacional de Luz S\'{i}ncrotron, Caixa Postal 6192, CEP 13083-970, Campinas, SP, Brazil}
\author{J.J. Neumeier}
\affiliation{Department of Physics, Montana State University, Bozeman, Montana 59717}
\author{J.W. Lynn}
\affiliation{NIST Center for Neutron Research, National Institute of Standards and Technology, Gaithersburg, Maryland 20899}
\affiliation{Center for Superconductivity Research, University of Maryland, College Park, Maryland 20742}
\author{D.N. Argyriou}
\affiliation{Materials Science Division, Argonne National Laboratory, Argonne, Illinois 60439}
\affiliation{Hahn-Meitner-Institut, Glienicker str. 100, Berlin D-14109, Germany}

\date{\today}

\begin{abstract}
A detailed investigation of mesoscopic magnetic and crystallographic phase separation in Ca$_{1-x}$La$_{x}$MnO$_{3}$, $0.00\leq x\leq 0.20$, is reported. Neutron powder diffraction and DC-magnetization techniques have been used to isolate the different roles played by electrons doped into the $e_{g}$ level as a function of their concentration $x$. The presence of multiple low-temperature magnetic and crystallographic phases within individual polycrystalline samples is argued to be an intrinsic feature of the system that follows from the shifting balance between competing FM and AFM interactions as a function of temperature. FM double-exchange interactions associated with doped $e_{g}$ electrons are favored over competing AFM interactions at higher temperatures, and couple more strongly with the lattice $via$ orbital polarization. These FM interactions thereby play a privileged role, even at low $e_{g}$ electron concentrations, by virtue of structural modifications induced above the AFM transition temperatures.
\end{abstract}

\pacs{61.12.Ld; 75.25.+z; 75.30.Kz; 75.70.Kw; 75.70.Pa}

\maketitle

\section{\label{introduction}Introduction}

The physical properties of mixed valent perovskite manganites such as Ca$_{1-x}$La$_{x}$MnO$_{3}$ are dominated by the strong coupling of charge-orbital and spin degrees of freedom. This results in a family of materials that shows pronounced physical property responses to chemical doping, temperature, pressure and magnetic field. Of these responses, colossal magnetoresistance (CMR) - a dramatic drop in resistivity in an applied magnetic field - at optimal doping $x\sim 0.7$ has attracted the most attention as it raises the possibility of applications such as data storage devices and sensors. 

\begin{figure}
\includegraphics[width=0.35\textwidth]{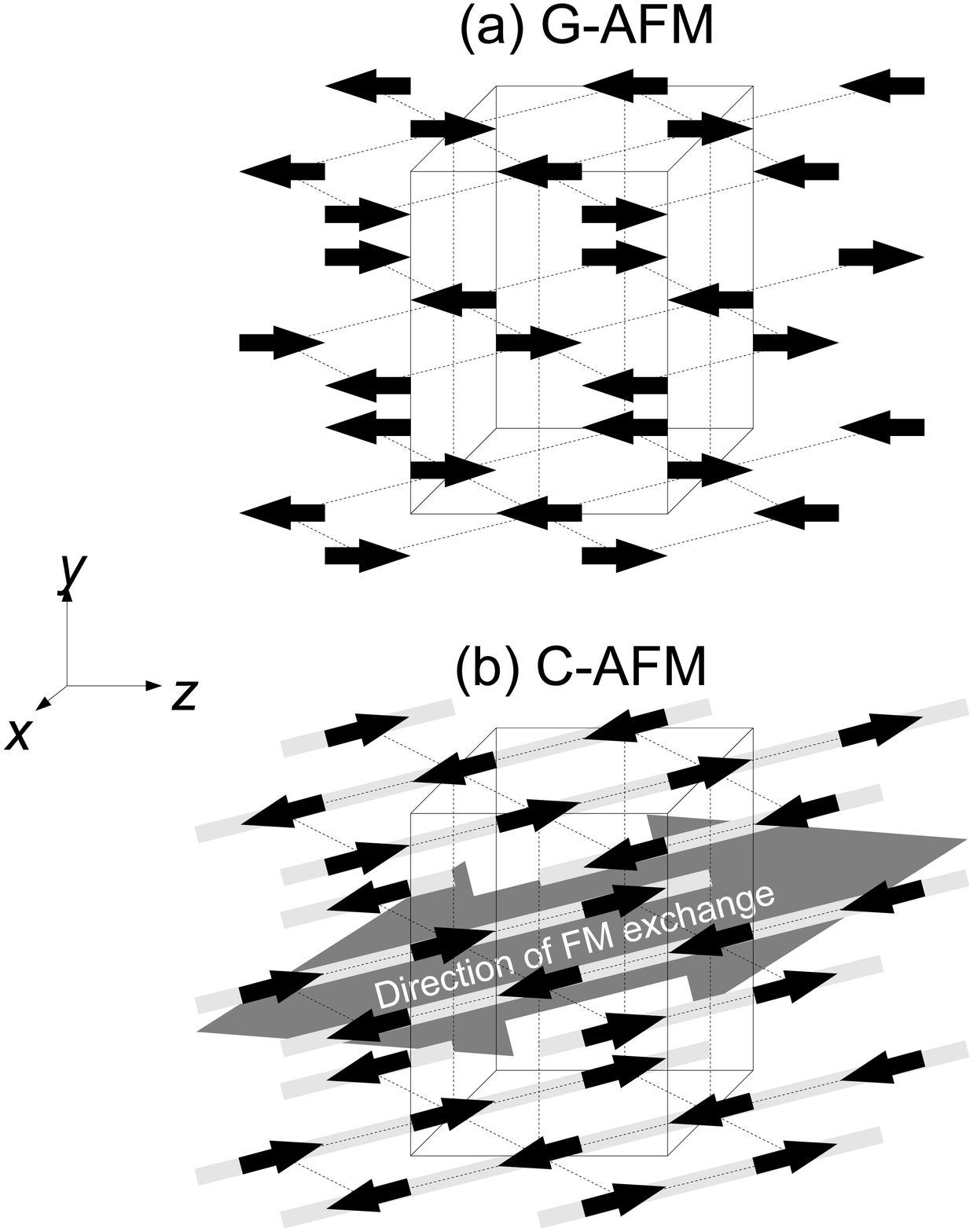}
\caption{\label{States}Schematic representations of the low-$T$ magnetic ground states of Ca$_{1-x}$La$_{x}$MnO$_{3}$ ($x\lesssim0.2$). Solid lines show the unit cell and dashed lines show nearest-neighbor Mn-Mn interactions. ($a$) Ideal G-AFM. ($b$) C-AFM, in which $e_{g}$ electrons delocalize into $d_{3z^{2}-y^{2}}$ orbital chains along the ($10\overline{1}$) direction, allowing 1-D FM DE while maintaining AFM interactions among the chains and causing a symmetry-lowering from orthorhombic P$nma$ to monoclinic P$2_{1}/m$.}
\end{figure}

At the \textquoteleft electron-doped\textquoteright\space end of the phase diagram ($x\sim 0$), the light doping of charges into the well understood G-type antiferromagnetic [G-AFM, Fig \ref{States}($a$)] ground state provides an opportunity to test the relevance of physical models of manganites. It was originally argued by de Gennes\cite{DeGennes60} that a small concentration of doped carriers into the \eg band (which is fully polarized due to the strong Hund's coupling of localized \tg electrons) gives rise to ferromagnetic (FM) double-exchange (DE) interactions, which for lightly doped systems competes with AFM super-exchange (SE) to produce a spin canted G-AFM state.\cite{Cheong99} This is consistent with a coexistence of FM and G-AFM components observed by neutron powder diffraction (NPD).\cite{Hagdorn99,Santhosh00,MartinJMMM99,Respaud01,Martin00} The application of simple DE across the whole of the phase diagram, however, contradicts experimental evidence by predicting either homogenous canting or the pure FM state at all points, in contrast with the rich phase diagram experimentally observed.\cite{Wollan55} In fact, the (ostensibly) spin-canted state only survives to $x\sim 0.1$, beyond which the degeneracy of the ($d_{x^{2}-y^{2}}$ and $d_{3z^{2}-r^{2}}$) \eg orbitals causes it to be supplanted by C-type AFM [C-AFM, Fig \ref{States}($b$)].\cite{Wollan55,Goodenough55,Santhosh00,Respaud01} In C-AFM, FM DE becomes long-range in one dimension $via$ delocalized $d_{3z^{2}-r^{2}}$ orbital chains, into which the doped \eg electrons are stabilized, while AFM SE is maintained perpendicular to these chains. This leads to a co-operative Jahn-Teller (J-T) distortion along the FM chain direction, lowering the symmetry from orthorhombic P$nma$ to monoclinic P$2_{1}/m$. 

\begin{figure}
\includegraphics[width=0.45\textwidth]{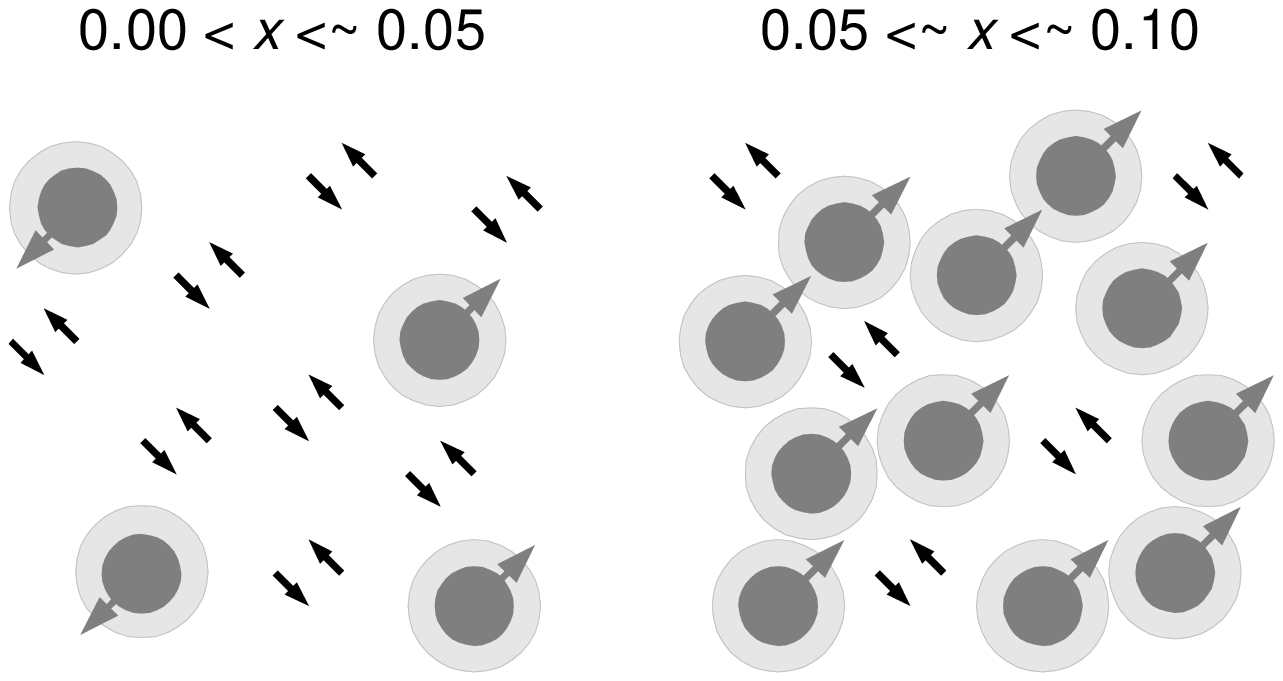}
\caption{\label{Clusters}Schematic representation of the growth of FM clusters within the G-AFM matrix (left), which can be aligned by an external applied magnetic field, into a FM cluster glass (right) exhibiting spontaneous long-range FM, with increasing $x$.\cite{GranadoXX} Black arrows represent the G-AFM matrix and grey spheres represent FM clusters.}
\end{figure}

These two papers report a detailed investigation into the nature of, and the relationships among, the rich variety of phases found in the electron-doped regime of Ca$_{1-x}$La$_{x}$MnO$_{3}$. This is of interest both as a model for spin-lattice coupling in the dilute limit of lattice polarons, and due to reports in many systems Ca$_{1-x}A_{x}$MnO$_{3}$ of large magnetoresistance effects \cite{Mahendiran00,Aliaga01,Chiba96,MaignanChemMat98,Martin99} and metamagnetic phase transitions.\cite{Filippov02,Respaud01} In Part I,\cite{GranadoXX} we found using neutron scattering that for light electron-doping $0.0<x\lesssim 0.1$, the G-AFM matrix contains a well-organized liquid distribution of FM clusters $\sim 10$ \AA\ in diameter. These clusters could be aligned by an external applied magnetic field to  produce a long-range FM moment, as seen at the opposite end of the same phase diagram.\cite{Hennion97,Hennion98,Hennion00} Higher density of these clusters at $x=0.09$ led to a spontaneous ($H=0$) long-range FM moment due to the formation of a FM cluster-glass,\cite{Maignan98} the orientation of which is coupled to the G-AFM matrix (Fig. \ref{Clusters}). Part II concerns principally the relationships among the various crystallographic (orthorhombic and monoclinic) and magnetic (G-AFM, liquid-like FM clusters, FM cluster glass and C-AFM) phases. High-resolution NPD and DC-magnetization techniques are used to address questions of sample homogeneity arising out of the observation of multiple crystallographic and magnetic phases in individual polycrystalline samples,\cite{Santhosh00,Mahendiran00,GranadoXX} necessary in order to correctly interpret local phenomenon observed using bulk probes. It is found that the inability to attain a unique thermodynamic ground state is an intrinsic feature of the system resulting from the extremely fine balance between competing states. 

\section{\label{experimental}Experimental Details}

Ceramic samples of Ca$_{1-x}$La$_{x}$MnO$_{3}$ with nominal compositions $0.00\le x\le0.20$ were prepared by solid state reaction. Stoichiometric quantities of ($99.99\,\%$ purity or better) CaCO$_{3}$, La$_{2}$O$_{3}$, and MnO$_{2}$ were weighed (to yield 7 g samples) and mixed in an agate mortar for 15 min followed by reaction for 20 h at 1100 $^{\circ}$C. The specimens were reground for 10 min, reacted for 20 h at $1150\,^{\circ}$C, reground for 10 min, reacted for 20 h at $1250\,^{\circ}$C, reground for 10 min, reacted for 46 h at $1300\,^{\circ}$C, reground for 10 min, reacted for 46 h at $1300\,^{\circ}$C, reground for 10 min, pressed into pellets, reacted for 17 h at $1300\,^{\circ}$C and cooled at $0.4\,^{\circ}$Cmin$^{-1}$ to $30\,^{\circ}$C. 

DC-magnetization measurements were conducted using a commercially available SQUID magnetometer.  Specimens were cooled to 5 K in zero field, then warmed to the highest measurement temperature in an applied field of $H=2000$ Oe. Magnetization $vs.$ $H$ curves were taken at 5 K. 

Temperature-dependent time-of-flight (TOF) NPD data were collected on the Special Environment Powder Diffractometer (SEPD) at Argonne National Laboratory's Intense Pulsed Neutron Source (IPNS). Data were analyzed by Rietveld-refinement using the program suite FullProf. 1-2 \% wt Marokite (CaMn$_{2}$O$_{4}$)\cite{Couffon64,Lepicard66} impurities were included in all refinements as both nuclear and (at low temperatures) magnetic phases (details of the low-$T$ AFM structure of Marokite are published elsewhere).\cite{Ling01} 

\section{\label{results}Results and Analysis}

\subsection{\label{DCanalysis}DC-magnetization and Resistivity}

\begin{figure}
\includegraphics[width=0.45\textwidth]{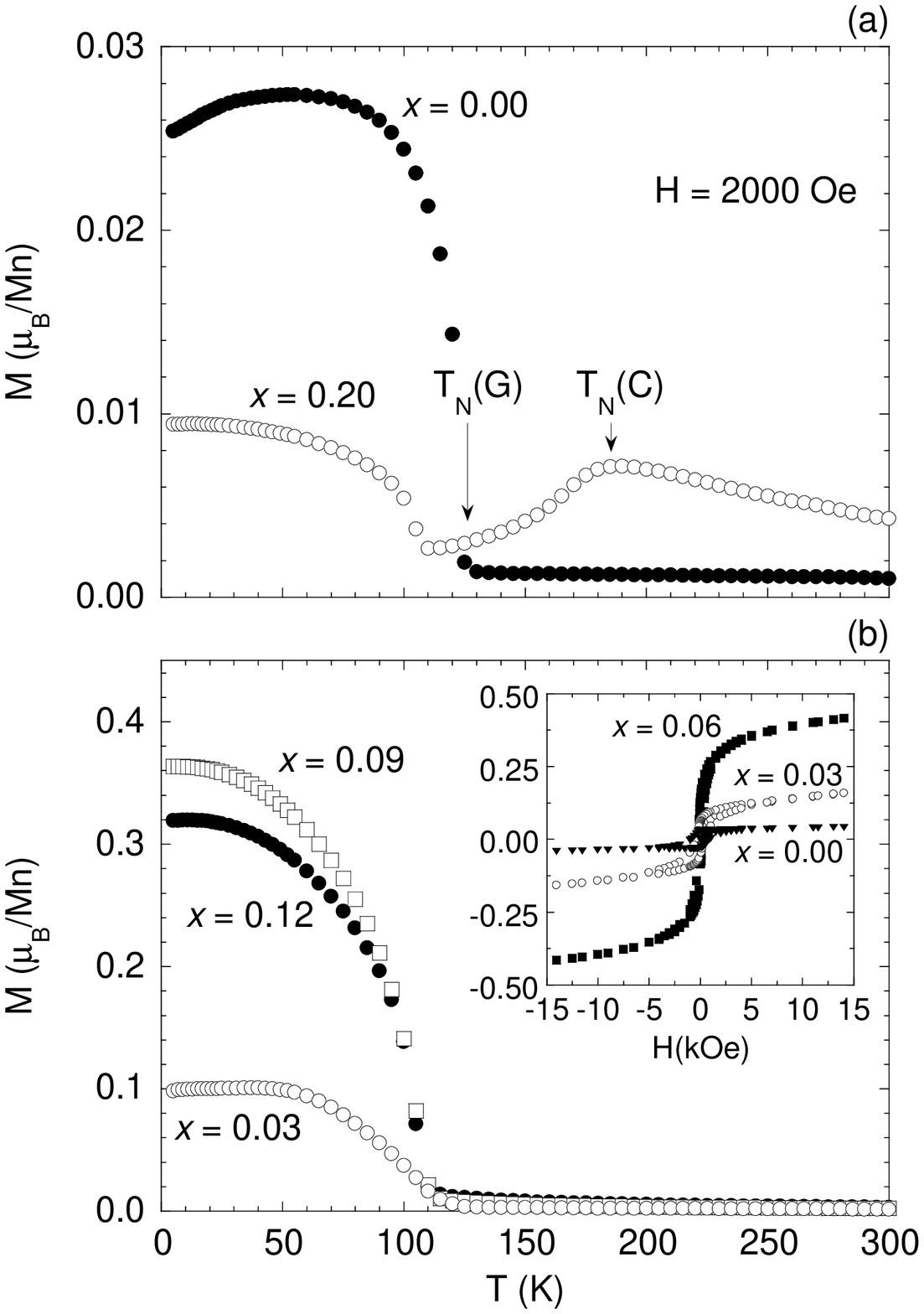}
\caption{\label{Magnetization}($a$) Magnetization $M$ $vs.$ $T$ measured at $H=2000$ Oe for $x=0.00$ and $x=0.20$ samples. $T_{N}$(C) and $T_{N}$(G) are indicated. ($b$) A similar plot (note the change in scale) for $x=0.03$, 0.09 and 0.12 samples, with an inset showing $M$ $vs.$ $H$ for $x=0.00$, 0.03 and 0.06. For the $M$ $vs.$ $H$ loops, lines were drawn through the low and high field regions of the data, and the intersection taken as the saturation moment $M_{sat}$ at 5 K used in Fig. \ref{M_vs_x}.}
\end{figure}

\begin{figure}
\includegraphics[width=0.45\textwidth]{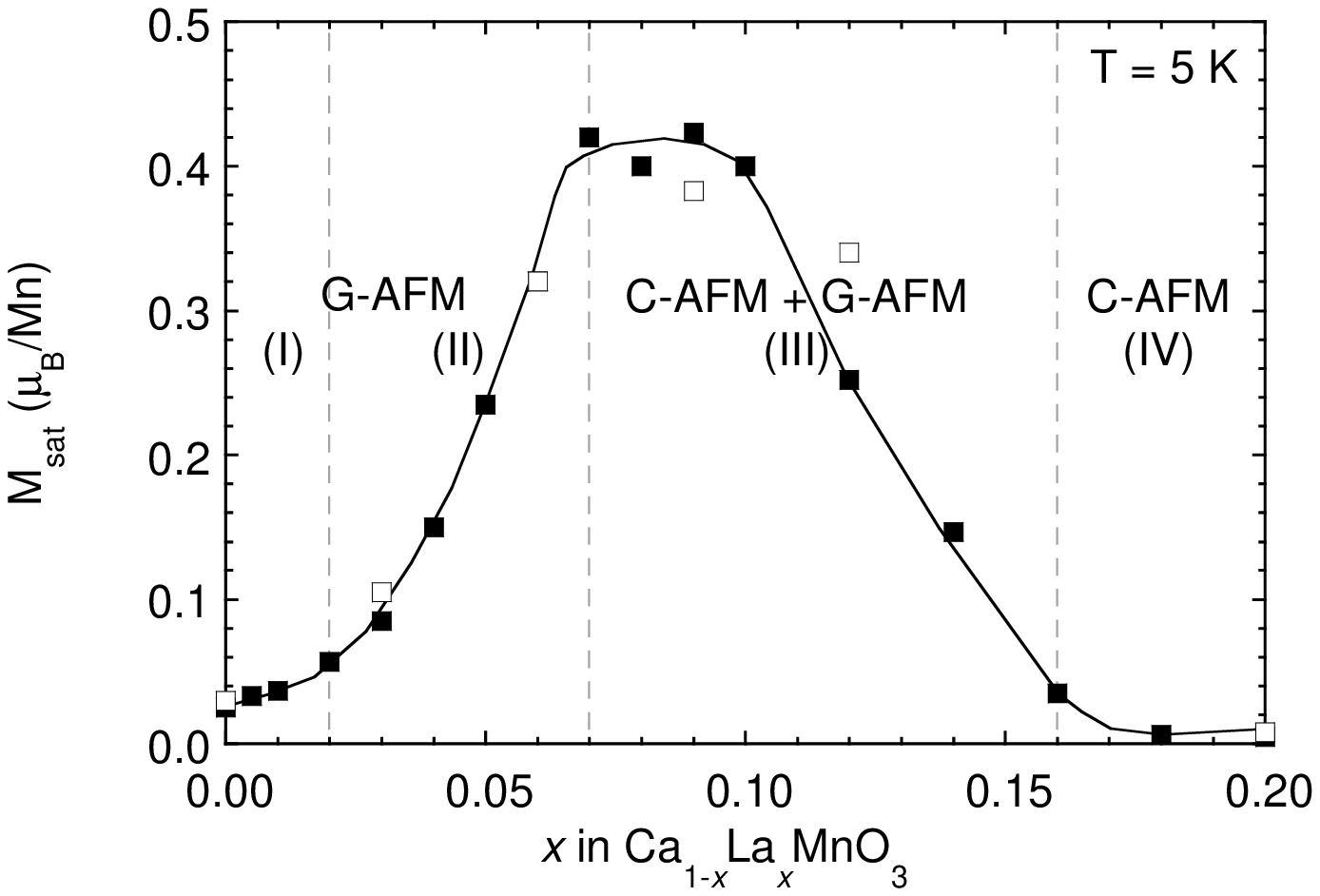}
\caption{\label{M_vs_x}Magnetic saturation moments $M_{sat}$ determined from $M$ $vs.$ $H$ curves at 5 K for the samples used in the present study (open squares) and in the study by Neumeier and Cohn \cite{Neumeier00} (solid squares) as a function of $x$. Regions (I-IV) discussed by Neumeier and Cohn are labelled and defined by dashed lines.}
\end{figure}

The $T$-dependencies of the DC-magnetization of all studied samples for $H=2000$ Oe are shown in Fig. \ref{Magnetization}. For $x=0.20$, a peak is observed at $T\sim 180$ K. A similar feature observed for a Ca$_{0.82}$Bi$_{0.18}$MnO$_{3}$ single-crystal \cite{Bao97}\@ was ascribed to a change of character of the spin fluctuations from FM to AFM with decreasing $T$, due to the freezing of the charge carriers and the consequent suppression of DE interactions. At lower $T$, a sudden enhancement in the DC-magnetization is observed below $T_{C}\sim 110-125$ K for $x\le 0.12$, and is ascribed to a spin-ordering transition with a FM component below $T_{C}$. The inset to Fig. \ref{Magnetization} shows DC-magnetization at 5 K as a function of $H$ for $x=0.00$, 0.03 and 0.06, with clear signatures of FM components (hysteresis). The $x$-dependence of the DC-magnetization at 5 K for $H=2000$ Oe is shown in Fig. \ref{M_vs_x}; results for the large samples used in this study (open markers) are consistent with those for smaller samples previously studied (closed markers).\cite{Neumeier00}

\subsection{\label{NPDanalysis}Neutron Powder Diffraction}

At 300 K, the crystal structures of samples at $x=0.00$, 0.03, 0.06,  0.09, 0.12, 0.16 and 0.20 were Rietveld-refined as single orthorhombic P$nma$ phases with TOF NPD data. In preliminary refinements, the O fractional occupancies [$\nu$(O)] were refined and found to lie between 0.98 and 1.02 for all samples, with typical standard deviations $\pm$0.008. This result supports the absence of significant cation or oxygen vacancies, consistent with chemical analysis performed on similarly prepared samples,\cite{Neumeier00} and $\nu$(O) was subsequently fixed at 1. 

\begin{figure*}
\includegraphics[width=0.62\textwidth,angle=-90]{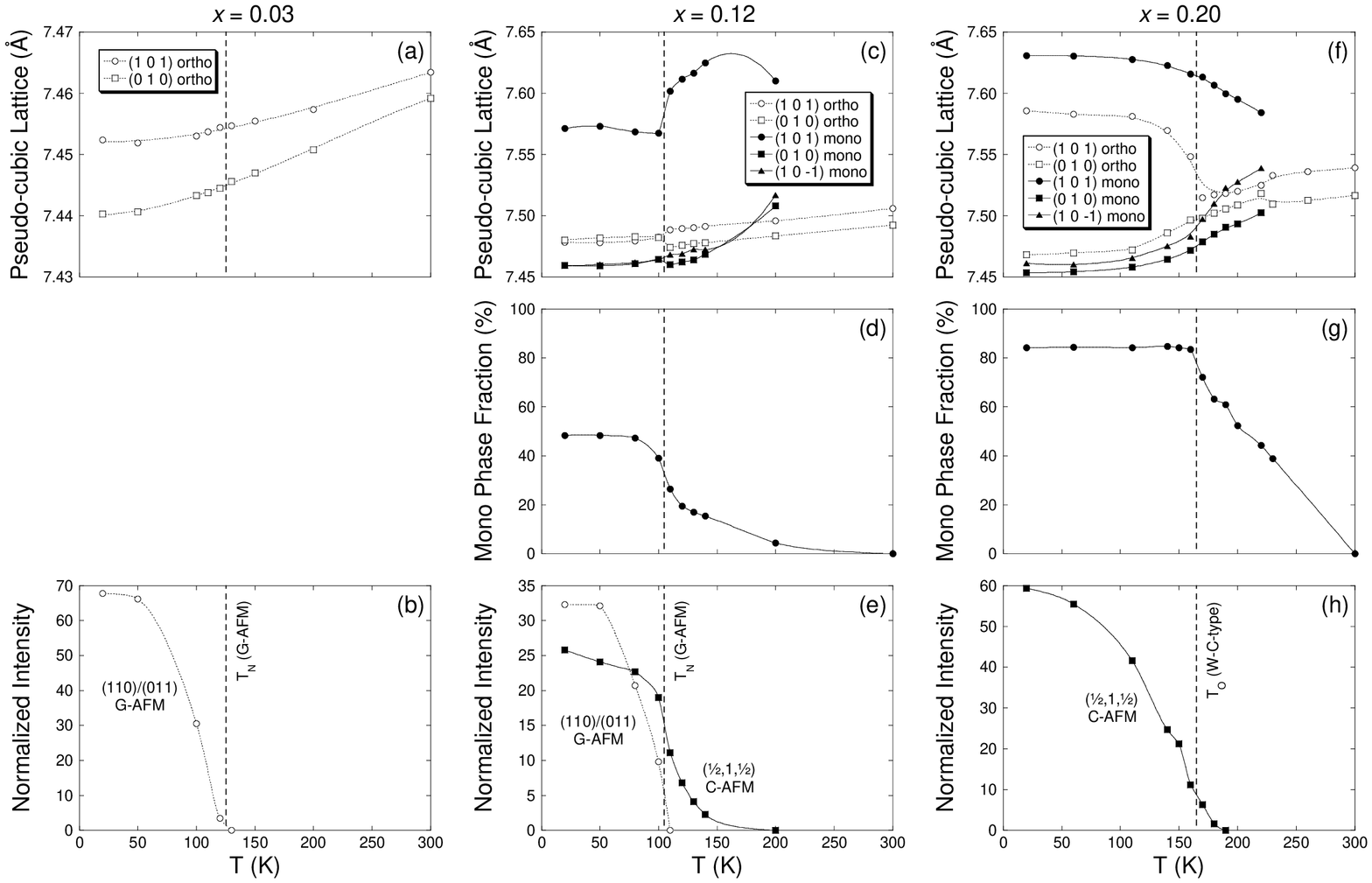}
\caption{\label{Lattice}($a$,$c$,$f$) $T$-dependence of pseudo-cubic lattice parameters ($101$), ($10\overline{1}$) and ($010$) for ($a$) $x=0.03$, ($c$) $x=0.12$ and ($f$) $x=0.20$, from Rietveld- refinement of NPD data. Note in ($c$) and ($f$) the elongation of the monoclinic unit cell along the ($10\overline{1}$) direction of $e_{g}$ electron polarization at $T_{N}$(C-AFM); and in ($f$) the elongation in the remaining orthorhombic phase of ($100$) and ($001$) due to the formation of a (twinned) Wigner-crystal type phase, in contrast to the isotropic monotonic contraction of the G-AFM orthorhombic cell in ($a$). ($d$,$g$) $T$-dependence of the phase fraction of symmetry-lowered monoclinic (P$2_{1}/m$) phase associated with the C-AFM magnetic phase, from the same Rietveld-refinement of NPD data. ($b$,$e$,$h$) $T$-dependence of the integrated intensities of the characteristic G-AFM (110)/(011) (open circles) and C-AFM ($\frac{1}{2}1\frac{1}{2}$) (filled circles) magnetic peaks from NPD data, normalized to the intensity of the strong nuclear (220)/(022) peak, for ($b$) $x=0.03$, ($e$) $x=0.12$ and ($h$) $x=0.20$.}
\end{figure*}

Figs. \ref{Lattice}($d$,$g$) show the crystallographic phase fraction of the symmetry-lowered (P$2_{1}/m$) phase associated with the C-AFM state, transformed from the orthorhombic P$nma$ state,  for $x=0.12$ and 0.20. This phase transition arises due to the polarization of $d_{3z^{2}-y^{2}}$ orbitals along the ($10\overline{1}$) direction, facilitating DE along the FM chains characteristic of C-AFM. A monoclinic phase fraction with a similar $T$-dependence could also be refined for $x=0.09$ and 0.16. (For $x=0.06$, although the presence of a weak C-AFM magnetic Bragg peak indicates the presence of a small monoclinic phase fraction, this fraction was too small to meaningfully refine.) The transition from the room-temperature orthorhombic P$nma$ phase in the C-AFM regime is also shown  in the plots of refined lattice parameters $vs.$ $T$ in Figs. \ref{Lattice}($c$,$f$). Note that for $x=0.20$ (Fig. \ref{Lattice}($f$), in addition to the monoclinic distortion undergone by the majority of the sample, the remaining orthorhombic phase fraction undergoes a different low-$T$ distortion, characterized by an elongation along ($100$) and ($001$). This type of distortion has been observed for Ca$_{2/3}$La$_{1/3}$MnO$_{3}$\cite{Radaelli99,Fernandez99} and Ca$_{1-x}$Bi$_{x}$MnO$_{3}$ ($x=0.22$, 0.25)\cite{Santhosh00} at low $T$, and has been ascribed to superstructures in the $ac$ plane caused by charge and orbital ordering of the Mn$^{3+}$ $e_{g}$ electrons (a \textquoteleft Wigner crystal\textquoteright-type or W-C-type phase). In order to account for the enlarged unit cell of W-C-type without excessively complicating the refinement, the O11 site was split evenly across the 4$f$ position. (Although the distortion clearly identifies this phase, no corresponding superstructure Bragg peaks were identified for our $x=0.20$ sample below the orbital-ordering temperature of the W-C phase $T_{O}$(W-C)$\sim 165$ K, possibly due to the small phase fraction and/or disorder.)

Final refined crystallographic phase fractions, unit cells, atomic positions, displacement parameters, Mn-O bond distances and magnetic moments at 20 K for $x=0.03$, 0.12 and 0.20 are given in Table \ref{Table1}. The differences between the Mn-O bond distances are very small for the orthorhombic G-AFM phase at $x=0.03$ and 0.12, $i.e.$ the MnO$_{6}$ octahedra are not significantly distorted. The same was true of the orthorhombic phases at $x=0.06$ and 0.09 at 20 K (not shown). Conversely, for the monoclinic C-AFM phase at $x=0.20$ and 0.12, as well the monoclinic phases at $x=0.16$, 0.09, and 0.06 (not shown), the Mn1-O12 and Mn2-O12 bonds are longer than the other Mn-O bond distances by $\sim 0.01-0.05$ \AA, $i.e.$ the MnO$_{6}$ octahedra are elongated along the ($10\overline{1}$) direction. This follows from the co-operative Jahn-Teller ordering along ($10\overline{1}$) that allows FM DE in the C-AFM state, and is consistent with previous diffraction studies for electron-doped CaMnO$_{3}$.\cite{MartinJMMM99,Pissas02} Finally, for the orthorhombic W-C-type phase at $x=0.20$, bond distances from Mn1 to the split O11 site illustrate the alternately shortened and elongated bonds along ($101$) and ($10\overline{1}$) characterizing this structure type and causing the elongation of $a$ and $c$ relative to $b/\sqrt{2}$ [see Fig \ref{Lattice}($f$)].

\begingroup
\squeezetable
\begin{table*}
\caption{\label{Table1}Results from Rietveld-refinements of TOF NPD data collected at 20 K for Ca$_{1-x}$La$_{x}$MnO$_{3}$ samples at selected values of $x$. Refinements were carried out in space groups P$nma$ (\#62) (atomic positions: O11 in $8d$, O21 in $4c$) and P$2_{1}/m$ (\#12) (atomic positions: O11 in $4f$, O12 in $4f$, O21 in $2e$, O22 in $2e$). G-AFM moments refined along ($001$), FM along ($010$) and C-AFM along ($10\overline{1}$). Superscript letters indicate constraints.}
\begin{ruledtabular}
\begin{tabular}{c d d d d d} 
$x$ & 0.03 & \multicolumn{2}{d}{0.12} & \multicolumn{2}{d}{0.20} \\
\hline \\
Phase fraction & 1 & 0.516(14) & 0.484(14) & 0.807(3) & 0.193(3) \\
Space group & \text{P}nma & \text{P}nma & \text{P}2_{1}/m & \text{P}2_{1}/m & \text{P}nma \\
$\mu$ ($\mu_{B}$/Mn) & \text{G-AFM\space} 2.47(3) & \text{G-AFM\space} 2.45(8) & \text{C-AFM\space} 2.29(7) & \text{C-AFM\space} 2.89(5) & - \\
& & \text{+ FM\space} 0.9(2) & & & \\
$a$ (\AA) & 5.27940(9)& 5.2933(5) & 5.3100(6) & 5.34495(17) & 5.3916(8)  \\
$b$ (\AA) & 7.44029(12) & 7.4731(6) & 7.4537(7) & 7.4617(2) & 7.4585(5) \\
$c$ (\AA) & 5.25978(8) & 5.2777(3) & 5.3218(8) & 5.33475(18) & 5.3561(2) \\
$\beta$ ($^{\circ}$) & - & - & 90.8457(18) & 91.3109(19) & - \\
LaCa1 $x$ & 0.0327(3) & 0.0323(14) & 0.030(3) & 0.019(2) & 0.0360(16) \\
LaCa1 $z$ & 0.9956(8) & 0.9964(19) & 0.989(3) & 0.004(2) & 0.9877(16) \\
LaCa2 $x$ & - & - & 0.524(3) & 0.523(2) & - \\
LaCa2 $z$ & - & - & 0.510(3) & 0.509(2) & - \\
O11 $x$ & 0.2860(2) & 0.2879(9) & 0.2799(17) & 0.2799(11) & 0.288(12)/ \\
& & & & & 0.271(5) \\
O11 $y$ & 0.03433(19) & 0.0314(5) & 0.0378(15) & 0.0402(10) & 0.022(6)/ \\
& & & & & 0.042(6) \\
O11 $z$ & 0.7128(3) & 0.7116(8) & 0.7162(17) & 0.7214(12) & 0.743(6)/ \\
& & & & & 0.699(6) \\
O12 $x$ & - & - & 0.7807(18) & 0.7839(6)^{a} & - \\
O12 $y$ & - & - & 0.0355(12) & 0.0294(8) & - \\
O12 $z$ & - & - & 0.7796(17) & 0.7839(6)^{a} & - \\
O21 $x$ & 0.4899(4) & 0.4866(12) & 0.495(3) & 0.4864(19) & 0.494(4) \\
O21 $z$ & 0.0678(5) & 0.0678(13) & 0.065(2) & 0.0639(18) & 0.053(4) \\
O22 $x$ & - & - & 0.998(2) & 0.9951(18) & - \\
O22 $z$ & - & - & 0.447(2) & 0.438(2) & - \\
Mn $B_{iso}$ & 0.20(3) & 0.22(7)^{d} & 0.22(7)^{d} & 0.20(6)^{b} & 0.20(6)^{b} \\
LaCa $B_{iso}$ & 0.39(2) & 0.35(7)^{e} & 0.35(7)^{e} & 0.36(6)^{c} & 0.36(6)^{c} \\
O11 $B_{iso}$ & 0.29(9) & 0.19(7)^{f} & 0.19(7)^{f} & 0.50(8) & 0.3(2) \\
O12 $B_{iso}$ & - & - & 0.19(7)^{f} & 0.29(7) & - \\
O21/22 $B_{iso}$ & 0.45(4) & 0.19(7)^{f} & 0.19(7)^{f} & 0.31(8) & 0.4(2) \\
Mn1-O11 (\AA) & 1.8970(13) & 1.904(5) & 1.888(9) & 1.908(6) & 2.03(2)/ \\
$\sim\parallel(101)$ & & & & & 1.83(3) \\
Mn1-O12 (\AA) & 1.9034(13) & 1.906(4) & 1.921(9) & 1.938(3) & - \\
$\sim\parallel(10\overline{1})$ & & & & & \\
Mn1-O22 (\AA) & 1.8947(5) & 1.9035(13) & 1.885(2) & 1.895(2) & 1.886(3) \\
$\sim\parallel(010)$ & & & & & \\
Mn2-O11 (\AA) & - & - & 1.916(9) & 1.898(6) & - \\
$\sim\parallel(101)$ & & & & & \\
Mn2-O12 (\AA) & - & - & 1.928(9) & 1.939(3) & - \\
$\sim\parallel(10\overline{1})$ & & & & & \\
Mn2-O21 (\AA) & - & - & 1.896(2) & 1.8980(18) & - \\
$\sim\parallel(010)$ & & & & & \\
$R_{B}$ & 0.0672 & \multicolumn{2}{d}{0.0612} & \multicolumn{2}{d}{0.0833} \\
$wR_{B}$ & 0.0700 & \multicolumn{2}{d}{0.0574} & \multicolumn{2}{d}{0.0843} \\
$\chi^{2}$ & 1.69 & \multicolumn{2}{d}{1.88} & \multicolumn{2}{d}{3.16} \\
\end{tabular}
\end{ruledtabular}
\end{table*}
\endgroup

\begin{figure}
\includegraphics[width=0.45\textwidth]{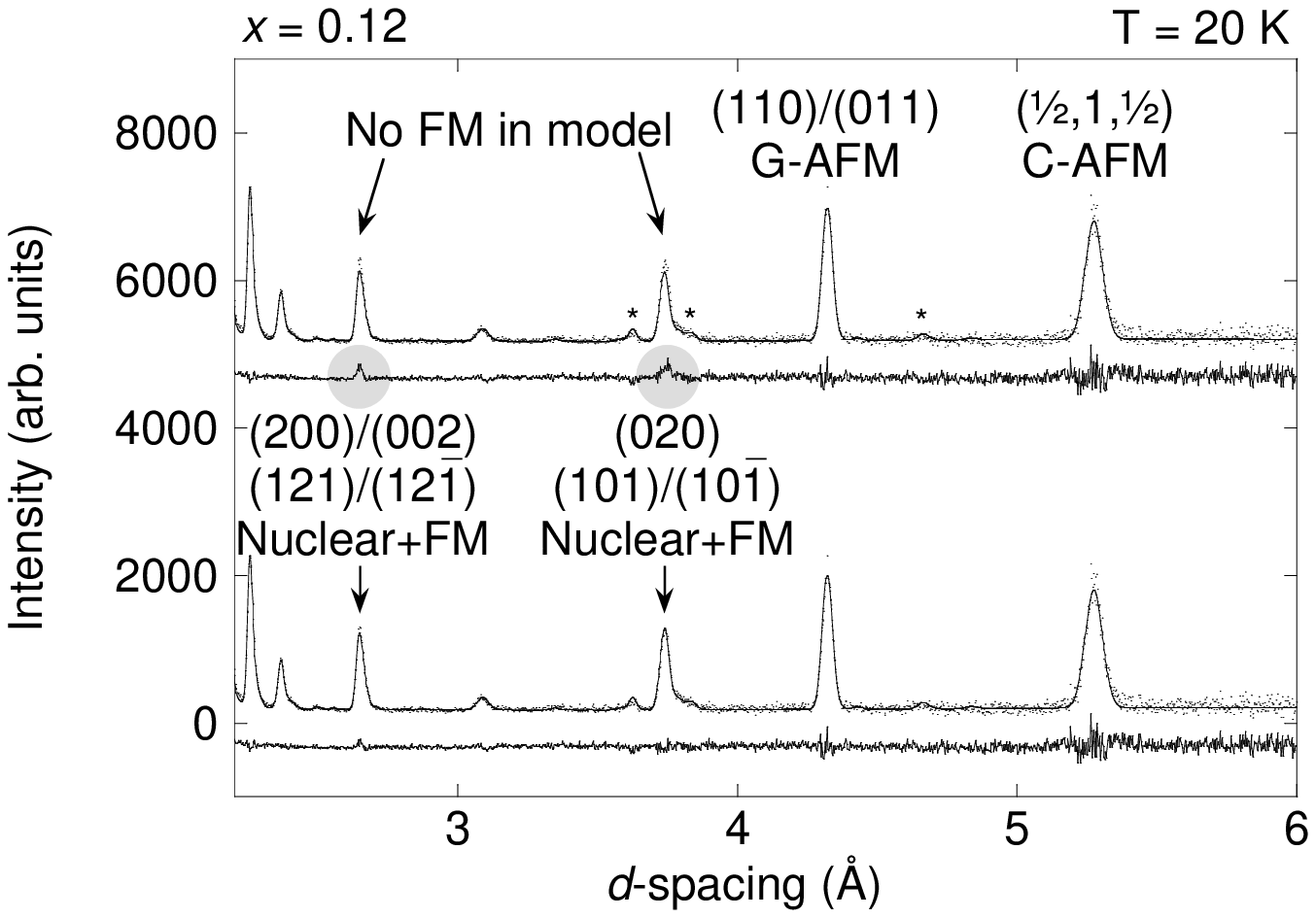}
\caption{\label{NPD}Observed (+), calculated and difference (below) plots of Rietveld-refined time-of-flight NPD data ($60\,^{\circ}$ detector bank of SEPD) for the $x=0.12$ sample at 20 K. Prominent magnetic reflections are labelled. The upper set shows the fit using a G-AFM + C-AFM magnetic model only, and the bottom set shows the fit with an additional FM component perpendicular to the G-AFM moment. Magnetic reflections due to the AFM Marokite impurity phase are marked (*).}
\end{figure}

The magnetic phases of the samples were also identified and Rietveld-refined by NPD, magnetic Bragg peaks being observed at low-$T$ for all samples. Observed magnetic reflections were consistent with G-AFM for $x=0.00$ and 0.03, and with C-AFM for $x=0.20$ and $x=0.16$. For intermediate dopings $x=0.06$, 0.09, and 0.12, G- and C-AFM Bragg reflections were observed simultaneously at low-$T$ (an extremely weak G-AFM peak was also observed for $x=0.16$). Fig. \ref{NPD} shows a portion of the TOF NPD pattern at high $d$-spacing (low-$Q$) at 20 K for $x=0.12$, illustrating the coexistence of Bragg peaks from distinct C- and G-AFM structures (dots); the solid lines in the upper set correspond to calculated and difference profiles using a magnetic model with phase-separated G- and C-AFM structures. Deficiencies in the fit shown in the upper set of Fig. \ref{NPD} are accounted for when a FM sublattice with spins perpendicular to those of the G-AFM lattice is included in the magnetic model (lower set). Significant FM intensities were also observed for $x=0.06$ and 0.09 (not shown). This is consistent with our DC-magnetization measurements, where the strongest FM signal was observed for $x$ between 0.06 and 0.12 (see Fig. \ref{M_vs_x}). No evidence was found for a FM moment in the monoclinic phase, as expected, all \eg electrons participating in FM DE along the ($10\overline{1}$) chain directions of C-AFM rather than forming FM clusters.

Figs. \ref{Lattice}($b$,$e$,$h$) show the $T$-dependencies of characteristic Bragg reflections associated with G- and C-AFM spin structures for $x=0.03$, 0.12, and 0.20 respectively. The magnetic ordering temperatures for G-AFM, $T_{N}$(G), correspond to the FM $T_{C}$ observed by DC-magnetization measurements (see Fig. \ref{Magnetization}) within experimental error. In contrast, the C-AFM order parameter for $x=0.12$, besides showing a rather peculiar $T$-dependence, is far more evident in the NPD than in the magnetization data. Note nonetheless that for $x=0.20$, $T_{N}$(C) $\sim 180$ K obtained by NPD is clearly associated with the peak in the DC-magnetization (compare Figs. \ref{Lattice}($h$) and \ref{Magnetization}($a$)). 

\section{\label{discussion}Discussion}

\subsection{\label{homogeneity}Sample Homogeneity}

The relationships among crystal lattices, crystallographic phase fractions and magnetic order parameters as functions of $T$ presented in Fig. \ref{Lattice} reveal relationships that bear on the compositional homogeneity of the polycrystalline samples used in this study. In particular, for $0.12$ and $0.20$ [Figs. \ref{Lattice}($d$,$g$)], the growth of the monoclinic phase fraction does not continue down to the lowest temperatures. For $x=0.12$ the phase fraction does not change significantly below $\sim T_{N}$(G-AFM), and for $x=0.20$ it does not change significantly below $T_{O}$(W-C). Furthermore, for $x=0.12$, $T_{N}$(G-AFM) (marked by the dashed line) appears to influence both the C-AFM magnetization [Fig. \ref{Lattice}($e$)] and the monoclinic lattice parameters [Fig. \ref{Lattice}($c$)]. These relationships imply that the different magnetic states are competing for the same domains within the sample, rather than simply forming in mutually exclusive, compositionally segregated, domains. 

The wide $x$-interval over which P$nma$ and P$2_{1}/m$ crystallographic phases coexist, which is typical of doped manganites $e.g.$ Ca$_{1-x}$Bi$_{x}$MnO$_{3}$\cite{Santhosh00} and Ca$_{1-x}$Sm$_{x}$MnO$_{3}$,\cite{Mahendiran00} cannot be understood in terms of mesoscopic inhomogeneities in composition $x$ within polycrystalline samples; rather, mesoscopic phase separation at low-$T$ is an intrinsic feature of electron-doped manganite perovskites, or at least of these systems where electron-doping is accomplished by compositional variation (phase separation might be favored by local chemical variations). The very recent study of Ca$_{1-x}$Sm$_{x}$MnO$_{3}$, $x=0.15$, by Algabarel \textit{et al.}\cite{Algarabel02} provides similar evidence for the existence of monoclinic C-AFM and orthorhombic FM-canted G-AFM in phase separated regions of compositionally homogeneous samples by demonstrating that their relative phase fractions could be influenced by an external applied magnetic field. There may be some compositional separation at low-$x$ because lighter doping gives a higher probability of La-clustering (as recently demonstrated in La$_{1-x}$Sr$_{x}$MnO$_{3}$\cite{Shibata02}\@), however, the significance of this decreases with increasing $x$. The competing magnetic states are extremely finely balanced over a broad crossover regime $0.06\le x\le 0.16$, within which samples do not settle into single thermodynamically stable phases at low-$T$. 

\subsection{\label{competition}Spin-lattice Coupling and Frustration}

The competition between orthorhombic G-AFM and monoclinic C-AFM [Figs. \ref{Lattice}($d$)] reflects the balance of gains and losses associated with the co-operative J-T distortion of the latter; a lowering in exchange energy on the one hand, and an increase in elastic energy on the other. This balance is affected by the relative strengths of FM DE and AFM SE interactions. The formation of the monoclinic phase below $T_{N}$(C-AFM) corresponds to the ordering of FM DE interactions (which exist as short-range fluctuations above $T_{N}$(C-AFM)\cite{Bao97}) into infinite 1-D chains, by AFM SE interactions perpendicular to them. The monoclinic phase fraction grows as $T$ decreases because these AFM SE interactions become stronger, decreasing the exchange energy of the monoclinic phase and making its total energy less than that of the paramagnetic orthorhombic phase. At $T_{N}$(G-AFM), however, the monoclinic phase fraction of the $x=0.12$ sample stops growing because AFM SE interactions become strong enough to stabilize G-AFM in the remaining orthorhombic phase fraction. This lowers the exchange energy and therefore the total energy of the orthorhombic phase, restoring its status as the more stable crystallographic polymorph. Similarly, in the $x=0.20$ sample [Figs. \ref{Lattice}($d$)], the monoclinic phase fraction stops growing at $T_{O}$(W-C) because the distortion required to form the W-C phase is less energetically costly than that required to form C-AFM. 

The effects of $T_{N}$(G-AFM) on the distortion of the pseudo-cubic lattices, as seen in Fig. \ref{Lattice}($a$ and $c$), are also of interest. The G-AFM structure is isotropic and therefore should not effect the lattice, as is indeed the case for $x=0.03$ [Fig. \ref{Lattice}($a$)]. For $x=0.12$, however, there seems to be a small but significant magnetostrictive effect at $T_{N}$(G-AFM) [Fig. \ref{Lattice}($c$)], whereby the ($010$) axis elongates slightly relative to ($101$). If this effect is real, it is presumably related to ($010$) being the direction of net FM in the cluster glass\cite{GranadoXX} (see Section \ref{NPDanalysis}); however, since the effect is small, speculation on a mechanism will be avoided. More surprising is the large effect of $T_{N}$(G-AFM) on the distortion of the monoclinic lattice [Fig. \ref{Lattice}($c$)], where no FM clusters are involved. Note firstly that the monoclinic phase does not adopt the fully-ordered C-AFM state immediately upon symmetry-lowering. This is clear for the $x=0.20$ sample, for which the magnetic order parameter [Fig. \ref{Lattice}($h$)] and monoclinic distortion [Fig. \ref{Lattice}($f$)] show a strong $T$-dependencies below $T_{O}$(W-C) [Fig. \ref{Lattice}($g$)], despite the fact that the monoclinic phase fraction no longer grows. In this light, the refined monoclinic cell for $x=0.12$ [Fig. \ref{Lattice}($c$)] might actually represent an \textit{average} monoclinic cell, for which a reduction in the monoclinic distortion would not necessarily represent a deterioration of established C-AFM ordered domains. It could simply be a convolution of the delay between symmetry-lowering and the establishment of a fully-ordered C-AFM state on the one hand, and the increasing strength of the competing G-AFM state on the other. 

An intriguing extension of this argument is the possibility that the increasing strength of AFM SE interactions as $T$ decreases not only slows the establishment of long-range FM DE ($i.e$ C-AFM) in the monoclinic phase, but actually leads to the establishment of G-AFM there instead. In the extreme case, G-AFM might actually replace established C-AFM domains. While there is no direct evidence for this in the present data, the reader's attention is brought to the highly analogous \textquoteleft bi-layered\textquoteright\space manganite perovskite La$_{2-2x}$Sr{$_{1+2x}$Mn$_{2}$O$_{7}$, where the C-AFM phase also requires a symmetry-lowering transition (from tetragonal to orthorhombic).\cite{Ling00} A 10 $\%$ electron-doped sample in this system ($x=0.90$) exhibited not only a structural transition followed by two magnetic transitions $T_{N}$(C-AFM) $=110$ K and $T_{N}$(G-AFM) $=60$ K, but a clear \emph{decrease} in the C-AFM order parameter below $T_{N}$(G-AFM); $i.e.$, the G-AFM state \textquoteleft colonizes\textquoteright\space the monoclinic regions established by C-AFM orbital polarization at higher-$T$. 

Low-$T$ phase inhomogeneities ultimately arise because short-range FM DE correlations appear at higher $T$ than AFM SE correlations, as has been noted in studies of weak diffuse neutron scattering above the magnetic long-range-ordering transition temperatures.\cite{Bao97,GranadoXX,Algarabel02} The C-AFM magnetic state can form at a higher temperature than the G-AFM state because the AFM interactions only have to be strong enough to create AFM order in 2-D, rather than 3-D. At the same time, the strength of the FM SE interactions is obviously related to the concentration of \eg electrons ($x$) facilitating it. Magnetic and crystallographic ground states are frustrated in the region where 1-D FM SE interactions are strong enough to cause \eg orbital polarization and symmetry-lowering at $T_{N}$(C), but where AFM DE is strong enough to create 3-D order below $T_{N}$(G-AFM). This frustration is illustrated by Fig \ref{PD}, where the phase diagram [Fig \ref{PD}($b$)] shows orthorhombic G-AFM to be the ground state for $x$ up to $0.16$, but Fig \ref{PD}($a$) shows that less than 20 \% of the $x=0.16$ sample is actually in this state at low-$T$.

\begin{figure}
\includegraphics[width=0.45\textwidth]{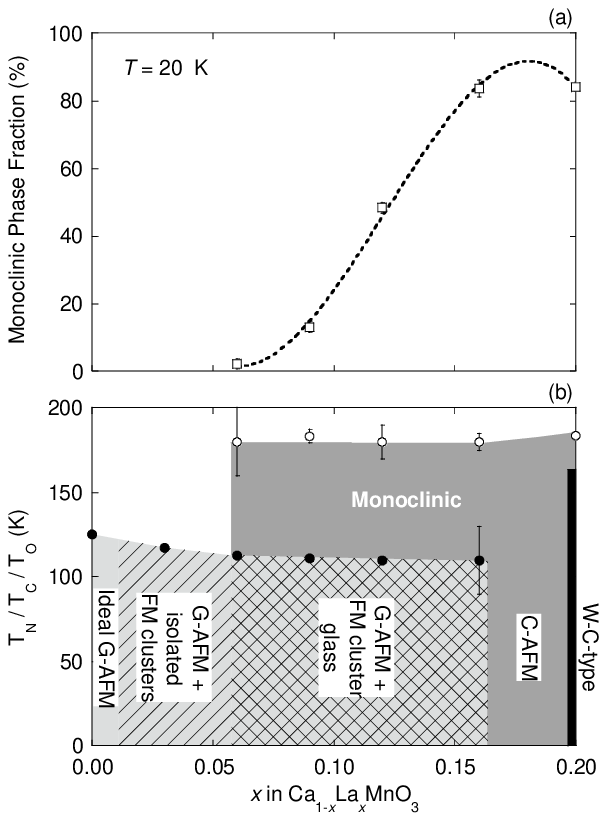}
\caption{\label{PD} ($a$) Refined monoclinic phase fraction at 20 K as a function is $x$. ($b$) Ground state phase diagram of Ca$_{1-x}$La$_{x}$MnO$_{3}$, $0.0\le x\le 0.2$, mapped onto the crystallographic and magnetic phase transitions determined from NPD data. Closed circles represent the coincident magnetic transitions $T_{N}$(G-AFM) and $T_{C}$, and open circles represent the coincident magnetic transition $T_{N}$(C-AFM) and structural transition $T_{O}$. Monoclinic (as opposed to orthorhombic) regions are dark grey. The presence of FM clusters is indicated by diagonal black lines, and the FM cluster glass by black diagonal cross-hatching. The W-C-type phase is black. Note that at the high-$x$ end of the G-AFM regime, the stability of the C-AFM state is very high, and therefore very little G-AFM is actually observed ($a$); the same is true for the W-C-type at $x=0.20$, and the converse is true at the low-$x$ end of the C-AFM regime.}
\end{figure}

\section{\label{conclusion}Conclusion}

This study (Part II) has used TOF NPD data in conjunction with physical property measurements to identify and characterize the low-$T$ phases present in samples of Ca$_{1-x}$La$_{x}$MnO$_{3}$ ($0.00\leq x\leq 0.20$) studied by neutron scattering in Part I.\cite{GranadoXX} The samples appear to be compositionally homogeneous and yet display multiple low-$T$ magnetic states, exemplifying the delicate balance among competing interactions characteristic of the CMR manganites. The results, in conjunction with those of Part I, have been used to construct a ground state phase diagram.

The theme that emerges from this phase diagram is thee strong effect that the introduction of FM DE interactions has on the AFM SE (for ideal G-AFM at $x=0$). In orthorhombic G-AFM these FM interactions create 0-D correlations; they have no influence on the long-range magnetic structure until the AFM SE interactions become strong enough below $T_{N}$(G-AFM) to create long-range 3-D AFM order, at which point they are \textquoteleft frozen in\textquoteright\space as isolated FM clusters, or (in sufficient densities) as an FM cluster-glass. In monoclinic C-AFM they become 1-D in character, leading to $d_{3z^{2}-r^{2}}$ orbital polarization and hence symmetry-lowering. At the low-$x$ end of the C-AFM regime, the actual thermodynamic ground state is G-AFM, but this is frustrated by the irreversible structural phase transition favoring C-AFM. In each case, FM DE interactions play a privileged role because they appear at higher-$T$ than the competing AFM SE interactions, allowing them to influence the structure on cooling and pre-dispose the system to a particular low-$T$ state. 

The changing dimensionality of the FM DE interactions  with $x$, from 0-D in the electron-doped G-AFM regime to 1-D in the C-AFM regime, foreshadows the subsequent change to 2-D (for the A-AFM state at $x\sim \frac{1}{2}$ in some manganite perovskite systems $e.g.$ Sr$_{1-x}$Pr$_{x}$MnO$_{3}$\cite{Martin99}) and finally 3-D (for the FM state at $\frac{1}{2}<x<1$ in most such systems). Between the fully-electron-doped (exhibiting a different type of A-AFM) and stoichiometric CaMnO$_{3}$ (G-AFM) end members, these changes in DE dimensionality with $e_{g}$ electron concentration underlie the magnetic phase diagram of the CMR manganites. At the same time, the ordering of these $e_{g}$ electrons $via$ the J-T effect underlies the crystallographic phase diagram. The C-AFM, A-AFM and FM magnetic states are examples of co-operation between these spin and orbital ordering effects, while the phase diagram is also punctuated by regions in which they compete, notably the C-E state at $x\sim 0.5$ and its W-C variants between the C-E and C-AFM regimes.

\begin{acknowledgments}
This work was supported by the U.S. Department of Energy, Basic Energy Sciences - Materials Sciences, under contract W-31-109-ENG-38 (CDL, DNA), by the National Science Foundation (NSF) under CAREER grant DMR 9982834 (JJN) and by FAPESP-Brazil (EG). Work at the university of Maryland was supported in part by the NSF MRSEC, DMR 00-80008. The authors thank S. Short of IPNS for technical support in the collection of NPD data.
\end{acknowledgments}

\bibliography{LCMO_PRB}

\end{document}